\documentclass[preprint]{aastex}
\usepackage{psfig}

\newcommand\mzon   {M$_{\odot}$}
\newcommand\pp     {$\pm$}

\def\degr{\hbox{$^\circ$}}
\newcommand\Lunit   {erg s$^{-1}$}
\newcommand\funit   {erg cm$^{-2}$ s$^{-1}$}
\def\degr{\hbox{$^\circ$}}

\righthead{{\itshape Chandra} observations of SAX J1747.0--2853}
\slugcomment{Accepted for publication in ApJ, 11 July, 2002}

\begin{document}

\title{{\itshape Chandra} observations of the bursting X-ray transient
SAX J1747.0--2853 during low-level accretion activity}

\author{Rudy Wijnands\altaffilmark{1,2}, Jon
M. Miller\altaffilmark{1}, Q. Daniel Wang\altaffilmark{3}}

\altaffiltext{1}{Center for Space Research, Massachusetts Institute of
Technology, 77 Massachusetts Avenue, Cambridge, MA 02139-4307, USA;
rudy@space.mit.edu}

\altaffiltext{2}{Chandra Fellow}

\altaffiltext{3}{Astronomy Department, University of Massachusetts,
Amherst, MA 01003, USA}

\begin{abstract}
We present {\it Chandra}/ACIS observations of the bursting X-ray
transient SAX J1747.0--2853 performed on 18 July 2001. We detected a
bright source at the position of R.A = 17$^h$ 47$^m$ 02.60$^s$ and
Dec. = --28\degr~52$'$ 58.9$''$ (J2000.0; with a 1$\sigma$ error of
$\sim$0.7 arcseconds), consistent with the {\it BeppoSAX} and {\it
ASCA} positions of SAX J1747.0--2853 and with the {\it Ariel V}
position of the transient GX +0.2,--0.2, which was active during the
1970's. The 0.5--10 keV luminosity of the source during our
observations was $\sim 3 \times10^{35}$ \Lunit~(assuming a distance of
9 kpc) demonstrating that the source was in a low-level accretion
state. We also report on the long-term light curve of the source as
observed with the all sky monitor aboard the {\it Rossi X-ray Timing
Explorer}. After the initial 1998 outburst, two more outbursts (in
2000 and 2001) were detected with peak luminosities about two orders
of magnitude larger than our {\it Chandra} luminosity. Our {\it
Chandra} observation falls in-between those two outbursts, making the
outburst history for SAX J1747.0--2853 complex. Those bright 2000 and
2001 outbursts combined with the likely extended period of low level
activity in-between those outbursts strongly suggest that the
classification of SAX J1747.0--2853 as a faint X-ray transient was
premature. It might be possible that the other faint X-ray transients
also can exhibit bright, extended outbursts which would eliminate the
need for a separate sub-class of X-ray transients.  We discuss our
results also in the context of the behavior of X-ray binaries
accreting at low levels with luminosities around $10^{35}$ \Lunit, a
poorly studied accretion rate regime.

\end{abstract}

\keywords{accretion, accretion disks --- stars: individual (SAX
J1747.0--2853) --- stars: neutron --- X-rays: stars}

\section{Introduction \label{section:intro}}

The X-ray transients are a class of X-ray binary systems which
occasionally exhibit bright outbursts during which they can display
X-ray luminosities of $10^{36}$ to $10^{39}$ \Lunit. Usually, those
sources stay this bright for only a few weeks to at most a few months
(see, e.g., Chen, Shrader, \& Livio 1997; Bradt et al. 2000), although
some can be active for several years. After the bright outbursts, the
X-ray transients slowly transit back into quiescence during which they
can only be detected in X-rays at a level of $10^{30}$ to $10^{34}$
\Lunit. The regime in-between quiescence and outburst ($10^{34}$ to
$10^{36}$ \Lunit) has not been studied extensively, partly due to the
relatively short time span the sources spent in this regime but also
due to the instrument limitations (e.g., lack of sensitivity or
observational constraints).

The neutron star sources which have been studied best in this regime
are Aql X-1 and the accretion-driven millisecond X-ray pulsar SAX
J1808.4--3658. Campana et al. (1998b) used {\it BeppoSAX} data to
study the decay of Aql X-1 after one of its outbursts. They found a
source which decreased steadily in luminosity until it reached
quiescent levels. Its spectrum hardened during the decay, although a
soft component below a few keV became more prominent. However, unlike
most X-ray transients, Aql X-1 also can exhibit long periods of
activity at a low level, which either occurs after a bright outbursts
or as a stand-alone event (Bradt et al. 2000; $\check{{\rm S}}$imon
2002). Such long episodes of low-level accretion have been observed
for only a few other systems, such as the black-hole system 4U
1630--47 (Kuulkers et al. 1997) or the neutron star systems 4U
1608--52 (Bradt et al. 2000; Wachter et al. 2002) and SAX
J1808.4--3658. This latter source was studied intensively by Wijnands
et al. (2001) using data obtained with the {\it Rossi X-ray Timing
Explorer} ({\it RXTE}) during the 2000 outburst of this source. They
found a source which was active at a low level (below a few times
$10^{35}$ \Lunit) for several months.  During this period, the source
exhibited violent variability on different time scales. Within only a
few days, the source decreased from a luminosity of a few times
$10^{35}$ \Lunit~to $\sim10^{32}$ \Lunit~(Wijnands et al. 2002;
Wijnands 2002), but a few days later it had increased again to
$10^{35}$ \Lunit. On several occasions, the source also exhibited
strong flaring behavior with a repetition frequency of $\sim1$ Hz (van
der Klis et al. 2000; Wijnands et al. 2000). The physical mechanisms
producing these violent fluctuations are not understood. Presently, it
is not known if other transients can display similar behavior at such
low luminosities, or if SAX J1808.4--3658 is unique in this sense
(e.g., due to its likely higher magnetic field strength compared to
the other transients).

Recently, a group of neutron star X-ray transients has been recognized
(e.g., Heise et al. 1999; in 't Zand 2001), which have a rather low
peak luminosity (a few times $10^{36}$ \Lunit) compared to the bright
transient sources (which have peak luminosities above $10^{37}$
\Lunit) and relatively short duration (e-folding time of less than a
week).  Observations of such dim transients are still sparse and it is
still unclear if they form a separate sub-class of transients (see the
discussion in In 't Zand 2001). King (2000) argued that they are
different from the bright systems and that they are neutron star X-ray
binaries which have evolved beyond their minimum orbital periods of
$\sim80$ minutes and have extremely low-mass companion stars.  The
above-mentioned millisecond X-ray pulsar is one of these dim X-ray
transients and is the best studied example of this possible sub-class
of transients.  Cumming, Zweibel, \& Bildsten (2001) suggested that
the low inferred time averaged mass accretion rate for SAX
J1808.4--3658 might be related to this source being a millisecond
X-ray pulsar.  For those systems which have high time averaged
accretion rates, the accreted matter will screen the magnetic field,
inhibiting the X-ray pulsation mechanism in those systems. This
suggests that the other dim neutron star X-ray transients, which
presumably have similar low time averaged accretion rate, might also
exhibit millisecond X-ray pulsations (e.g., Cumming et al. 2001).

Besides those 'faint X-ray transients', another, different group of
neutron-star X-ray binaries also have been identified which members
were only seen during type-I X-ray bursts and with an accretion
luminosity typically below $10^{36}$ \Lunit, which was undetectable by
the instruments used.  Cocchi et al. (2001) discussed the possibility
that those systems are a separate subclass of neutron star systems
which have persistent luminosities of order $10^{34-35}$
\Lunit~(called the 'low-persistent bursters'). This idea was supported
by the detection of the possible group members 1RXS J171824.2--402934
(Kaptein et al. 2000) and SAX J1828.5--1037 (Cornelisse et al. 2002a)
at a level of $\sim10^{34-35}$ \Lunit. Cornelisse et al. (2002a)
elaborated on the possibility of this extra sub-class of X-ray
bursters and found that their spatial distribution is consistent with
the general X-ray burster distribution, but different from that of the
above mentioned faint X-ray transients. This suggests that the faint
X-ray transients are a different source population than the
low-persistent bursters and the bright neutron star X-ray transients.
However, {\it Chandra} follow-up observations of several potential
group members showed that those sources could only be seen at a
luminosity below a few times $10^{32}$ \Lunit~(Cornelisse et
al. 2002b), which are similar to the quiescent luminosities observed
for the bright neutron star X-ray transients. This suggests that those
low-luminosity persistent sources might be genuine transients but have
sub-luminous ($10^{34-35}$ \Lunit) outbursts episodes, possible with a
durations of years (see the discussion in Cornelisse et
al. 2002b). Clearly, the behavior of X-ray binaries at luminosities
below $10^{36}$ \Lunit~ is rather complex and our knowledge of the
basic observational properties is very limited.

In 1998 March, a new X-ray transients was discovered by In 't Zand et
al. (1998) using observations of the Galactic center region with the
Wide Field Cameras (WFC) aboard {\it BeppoSAX}. The position of this
new source (designated SAX J1747.0--2853) is consistent with that of
the transient source GX +0.2,--0.2, which was observed in the 1970's
(Proctor, Skinner, Willmore 1978). From the detection of type-I X-ray
bursts (In 't Zand et al. 1998) is it clear that the compact object in
this system is a neutron star because such events are thought to be
due to thermonuclear flashes on the surface of such a star. The source
was observed with the Narrow Field Instruments (NFI) aboard {\it
BeppoSAX} on several occasions. Using these NFI observations, both
Sidoli et al. (1998) and Natalucci et al. (2000) found a source whose
spectrum was consistent with that of the neutron star X-ray binaries
when they are at a luminosity of approximately a few times $10^{36}$
\Lunit~(e.g., Barret et al. 2000), similar to the luminosity observed
for SAX J1747.0--2853.  From the observed type-I bursts a distance of
$\sim9$ kpc was obtained (Natalucci et al. 2000). The last reported
detection of the source during this outburst was on 1998 April 15
(Sidoli et al. 1998) after which the source is presumed to have become
quiescent again.  The low peak luminosity of the source spurred the
tentative classification of the source as a faint X-ray transient
(Heise et al. 1999).

Markwardt et al. (2000a) reported that in early March 2000, SAX
J1747.0--2853 could be detected again using the proportional counter
array aboard the {\it RXTE} satellite. During this outburst, the
source was also detected by {\it ASCA} (Murakami et al. 2000) and {\it
BeppoSAX} (Campana, Israel, \& Stella 2000). Here we present {\it
Chandra} observations performed in July 2001 of the region including
SAX J1747.0--2853.

\section{Observation, analysis, and results}

SAX J1747.0--2853 was in the field of view of two observations
performed as part of {\it Chandra} Galactic Center Survey (GCS; Wang,
Gotthelf, \& Lang 2002): observations GCS 10 and GCS 11 (see
Tab.~\ref{tab:log} for more details about those observations). The
source was approximately 2.7$'$ and 14.3$'$ located from the nominal
pointing direction during respectively the GCS 10 and GCS 11
observation.  The ACIS-I instrument was used during these observations
in combination with the S2 and S3 chips of the ACIS-S instrument.  To
limit the telemetry rate only those photons with energy above 1 keV
were transmitted to Earth. The data were analysed using the analysis
package CIAO, version 2.2.1, and the threads listed on the CIAO web
pages\footnote{Available at http://asc.harvard.edu/ciao/}.

\subsection{The image and position of SAX J1747.0--2853}

\begin{figure}
\begin{center}
\begin{tabular}{c}
\psfig{figure=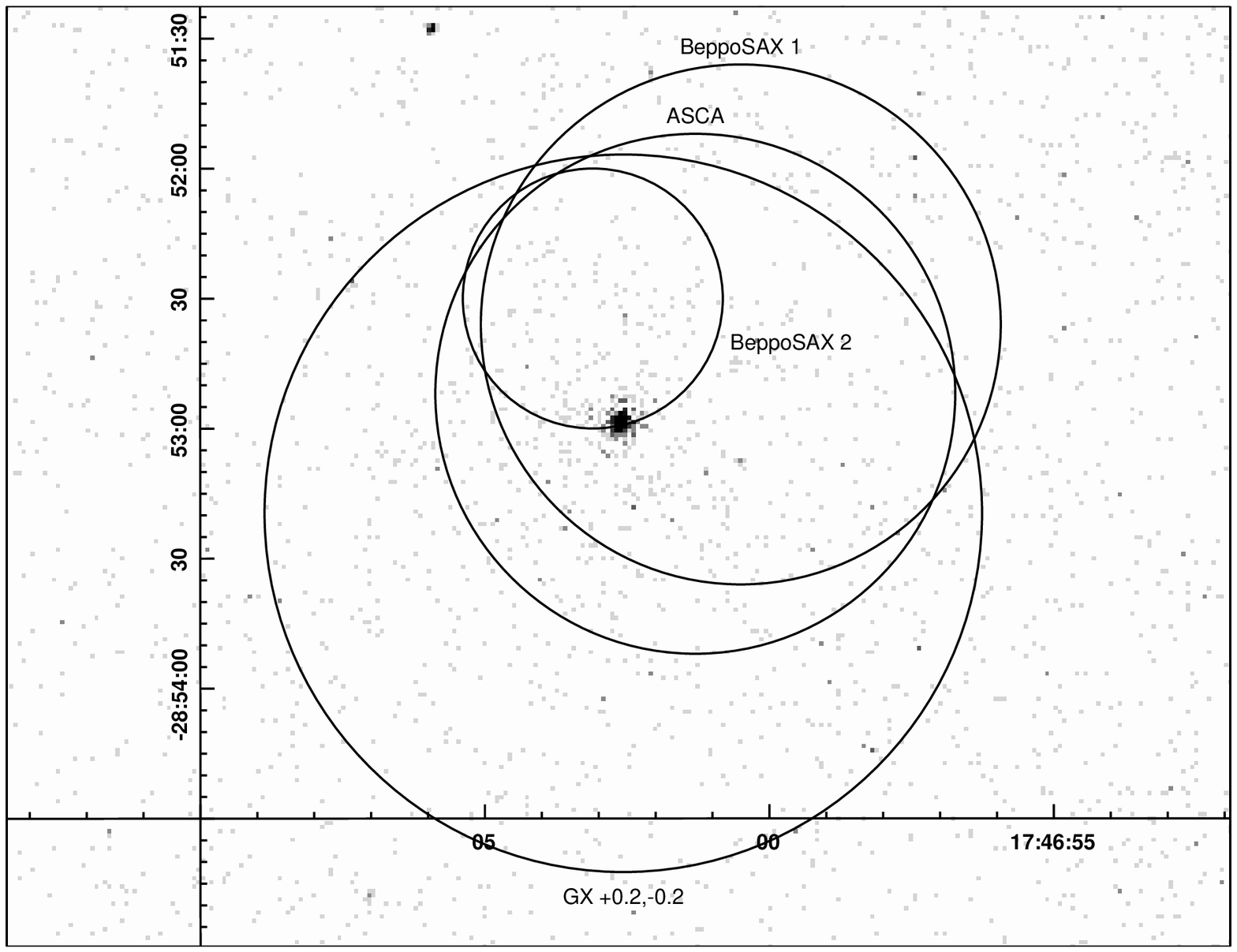,width=16cm}
\end{tabular}
\figcaption{The {\it Chandra}/ACIS-I image (for photons with energy of
1 keV or higher) of SAX J1747.0--2853 during observation GCS 10. The
coordinates are for epoch J2000.0.  Shown are the {\it BeppoSAX} (1:
Sidoli et al. 1998; 2: Campana et al. 2000) and the {\it ASCA}
(Murakami et al. 2000) error circles on the position of SAX
J1747.0--2853. Furthermore, the {\it Ariel V} error circle of GX
+0.2,--0.2 (Proctor et al. 1978) is also displayed. The detected
source is consistent with all positions of SAX J1747.0--2853 and with
the position of GX +0.2,--0.2. \label{fig:image} }
\end{center}
\end{figure}

In Figure~\ref{fig:image}, we show the region containing SAX
J1747.0--2853 as observed during observation GCS 10. We clearly detect
a bright source in the various error-circles of SAX J1747.0--2853
(although at the edge of the error-circle given by Campana et
al. 2000) and in that of GX +0.2,--0.2. This source was also present
during the GCS 11 observation (not shown), however, the source was
located at the edge of the ACIS-S2 chip and due to the large point
spread function (PSF) for sources which are $\sim14'$ off-axis the
source appears very extended and a considerable fraction of the source
photons do not fall on the chip. The positional coincidence of our
detected source with SAX J1747.0--2853 and GX +0.2,--0.2 makes it
likely that our source can be identified with those two transients.

We used the CIAO tool WAVDETECT to obtain the coordinates of SAX
J1747.0--2853: R.A = 17$^h$ 47$^m$ 02.604$^s$ and Dec. =
--28\degr~52$'$ 58.9$''$ (J2000.0; with a 1$\sigma$ error of $\sim$0.7
arcseconds; Aldcroft et al. 2000). Although we detected many other
sources beside SAX J1747.0--2853, none of them were located in the
various error-circles of SAX J1747.0--2853 and GX +0.2,--0.2. Also,
none of those near-by sources could conclusively be identified with a
star in either the USNO2 (Monet et al. 1996) or the second incremental
data release of the 2 Micron All Sky Survey (2MASS)
catalog. Therefore, we could not improve on our {\it Chandra} position
of SAX J1747.0--2853.  No USNO2 or 2MASS star is visible at the
position of SAX J1747.0--2583 (Fig.~\ref{fig:finding}), which is not
surprising because at the time of those near-infrared observations
(1998 July 5), the source was presumed to be in quiescence.  In
Figure~\ref{fig:finding} we present the $J$ and $K_{\rm s}$ finding
charts of SAX J1747.0--2853 obtained from the 2MASS archive.

During the GCS 10 observation, we detected 1740\pp42 counts
(background corrected) from the source, resulting in a count rate of
0.147\pp0.004 counts s$^{-1}$. This count rate has not be corrected
for pile-up, which is a serious effect for a source this bright (see
also section~\ref{section:spectrum}). The exact count rate during the
GCS 11 observation is difficult to estimate because the source falls
only partly on the chip\footnote{To obtain an accurate correction
factor for the number of counts detected and the observed flux of the
source, one has to know accurately the PSF for a source this far
off-axis, and the exact dither pattern used in order to estimate the
total exposure of the source on the chip. Such detailed estimates are
beyond the scope of this paper.\label{footnote}}; the number of counts
is $>757$ counts resulting in a count rate of $>0.065$ counts
s$^{-1}$.  SAX J1747.0--2853 is known to exhibit type-I X-ray bursts
(In 't Zand et al. 1998; Sidoli et al. 1998; Natalucci et al. 2000)
and, therefore, we made source light curves for both observations. We
did not observe any X-ray bursts.

\subsection{The X-ray spectrum of SAX J1747.0--2853 \label{section:spectrum}}

We have extracted the source spectrum using a circle with a radius of
15$''$ on the source position. The {\it Chandra} PSF combined with the
brightness of the source made such a large extraction region necessary
to encompass most of the source photons. The background data were
obtained by using an annulus on the same position with an inner radius
of 30$''$ and an outer one of 60$''$. This region was chosen in order
to obtain a background which did not contain any source photons. The
data were rebinned using the FTOOLS routine grppha into bins with a
minimum of 15 counts per bin.

We started out by fitting the obtained spectrum with XSPEC (version
11.1; Arnaud 1996). However, when fitting a power-law model to the
data, we obtained a low photon index of $\sim1$ and a low column
density of $\sim4\times10^{22}$ cm$^{-2}$, which was inconsistent with
that observed during the {\it BeppoSAX}/NFI observations of SAX
J1747.0--2853 (Sidoli et al. 1998; Natalucci et al. 2000).  This
strongly suggest that the spectrum is heavily piled-up, resulting in
an artificially hard X-ray spectrum. In order to correct for the
pile-up, we used the analysis package ISIS, version 0.9.81 (Houck \&
DeNicola 2000), and the pile-up model available there in (Davis
2001). The resulting fit parameters are listed in
Table~\ref{tab:spectrum}.

To check our results we have also extracted the spectrum of the source
using an annulus around the source position of 1.5$''$ to 15$''$, thus
excluding the data which were heavily affected by pile-up. The
obtained spectral parameters are in good agreement with those obtained
using ISIS (see Tab.~\ref{tab:spectrum}; note that the errors on the
parameters are large due to the much lower number of photons in the
spectrum). In our extraction annulus only $\sim$5\% of the source
photons are located; we have corrected our obtained flux for this, and
this flux is consistent with that we have obtained from fitting the
piled-up spectrum in ISIS.  We also extracted the source spectrum from
observation GCS 11 and fitted it in XSPEC. The extended nature of the
source during this observation (due to the large PSF) makes it
unlikely that the source suffers from a significant amount of pile-up
and no pile-up correction was applied.  Again, we obtained similar
spectral parameters (Tab.~\ref{tab:spectrum}). No flux is listed
because the correction factor is unknown (see
footnote~\ref{footnote}).

We used PIMMS\footnote{A web version is available at
http://heasarc.gsfc.nasa.gov/Tools/w3pimms.html} in order to estimate
the pile-up fraction in our spectrum. Using the parameter range found
for our GCS 10 spectrum, we obtained a pile-up fraction between 32 and
47 \% and a count rate after pile-up of 0.146 to 0.161 counts
s$^{-1}$, which is consistent with our detected count rate of
0.147\pp0.004 counts s$^{-1}$.

We have also fitted our spectrum with different spectral models, such
as a black-body model. This model resulted in a column density of
3.8$^{+0.8}_{-0.7}$ or 4.4$^{+1.0}_{-0.8} \times 10^{22}$ cm$^{-2}$
and a temperature $kT$ of 1.5\pp0.2 or 1.2\pp0.2 keV, for observation
GCS 10 (using the pile-up model in ISIS) or GCS 11, respectively. The
flux obtained for observation GCS 10 is $1.1\times10^{-11}$
\funit~(0.5--10 keV; unabsorbed). However, this model gave a rather
low column density compared to what has been measured before with {\it
BeppoSAX} (Sidoli et al. 1998; Natalucci et al. 2000), indicating that
this might not be the correct model to fit the data. A cut-off
power-law, a thermal comptonization model, or a bremsstrahlung model
could also be used to fit the data accurately, however, for simplicity
we only list the results of the power-law model.

\subsection{Long-term behavior}

To investigate the long-term behavior of the source, we have plotted
the {\it RXTE} all sky monitor (ASM; Levine et al. 1996; available at
http://xte.mit.edu/ASM\_lc.html) light curve in
Figure~\ref{fig:asm_lc}.  The source was clearly detected during the
2000 outburst (e.g., Markwardt et al. 2000a) and it again showed a
re-flare in the summer/fall of 2001.  The time of our {\it Chandra}
observations is indicated in the figure by the arrow. At the time of
these observations the ASM data seem to indicate that SAX
J1747.0--2853 was detected by this instrument. However, using PIMMS
and the spectral results from our {\it Chandra} data, the predicted
count rate for the ASM is $\sim0.04$ counts s$^{-1}$, which is below
the sensitivity of the ASM. This strongly suggest that the activity
detected by the ASM in-between the two outbursts is unrelated to SAX
J1747.0--2583 and probably it is due to the uncertainties in the
fitting process (Levine et al. 1996) and the proximity of SAX
J1747.0--2853 near the Galactic center. This is consistent with the
low fluxes detected for this source during the Galactic bulge scan
program (Markwardt et al. 2000b) of {\it RXTE} as judged from the
light curve of SAX J1747.0--2853, which is presented by In 't Zand
(2001) in his Figure 4.

Assuming a power-law spectrum with index of 2 and a column density of
$7\times10^{22}$ cm$^{-1}$, 1 ASM count corresponds to an unabsorbed
0.5--10 keV flux of $9.5\times10^{-10}$ \funit. The peak count rate of
$\sim8$ and $\sim4$ counts s$^{-1}$ during the 2000 and 2001 outburst
would results in a flux of $7.6$ and $3.8\times10^{-9}$ \funit,
respectively. These fluxes are obtained using 5 day averaged count
rates and the ASM dwell data indicates that at certain times the
fluxes could be a factor of 2 higher. Furthermore, extra uncertainties
are introduced by our assumption of a power-law spectrum for the
source, but usually neutron star X-ray transients become softer when
they increase in luminosities. Despite these uncertainties, these
fluxes are clearly considerably larger (up to two orders of magnitude)
than the fluxes we observed during our {\it Chandra} observations.

\section{Discussion\label{section:discussion}}

We have detected a bright X-ray source in the various error circles of
SAX J1747.0--2853 and in that of GX +0.2,--0.2, making it likely that
our source can be identified with those transients. The source had a
0.5--10 keV unabsorbed X-ray flux of $\sim3.3 \times 10^{-11}$ \funit,
which for a distance of 9 kpc (Natalucci et al. 2000) results in a
0.5--10 keV luminosity of $\sim3\times10^{35}$ \Lunit.  This
luminosity is about two orders of magnitude lower than the maximum
luminosity observed during the peak of the 2000 and 2001 outbursts (as
observed with the {\it RXTE}/ASM). These high outburst luminosities
are different from the low (a few times $10^{36}$ \Lunit; 2--10 keV)
luminosity observed for this source during its 1998 outburst (Sidoli
et al. 1998; Natalucci et al. 2000). SAX J1747.0--2853 is in this
respect similar to several other systems, such as for example the
neutron star system Aql X-1 or the black-hole systems XTE J1550--564
and XTE J1859+226.  Aql X-1 usually exhibits bright outbursts, but
occasionally it exhibits outburst which are one order of magnitude
less bright (e.g., Bradt et al. 2000; $\check{{\rm S}}$imon 2002). XTE
J1550--564 exhibited a bright outburst in 1998/1999 (see, e.g.,
Sobczak et al. 2000) but after that three much weaker outbursts have
been observed (Smith et al. 2000; Tomsick et al. 2001a; Swank, Smith,
\& Markwardt 2002). Similarly, XTE J1859+226 was detected as a bright
transient in October 1999 (Wood et al. 1999; Markwardt, Marshall, \&
Swank 1999), but later it was found to exhibit a small re-flare
(Casares et al. 2000; Miller et al. 2000). Clearly, X-ray transients
can exhibit a variety of outbursts profiles and can have very bright
outbursts followed by very weak ones, or vice versa.

\subsection{Faint neutron-star X-ray transients}

Recently, a group of faint neutron star X-ray transients was
recognized (e.g., Heise et al. 1999; in 't Zand 2001), with outburst
peak luminosities of only $10^{36-37}$ \Lunit~ and short e-folding
time scales (less than a week), implying a time-averaged accretion
rate of only $\sim10^{-11}$ \mzon~ yr$^{-1}$.  Based on the
characteristics of the 1998 outburst of SAX J1747.0--2853, this source
was put into this class of faint transients.  However, the later
outbursts (especially that in 2000) demonstrate that this source can
also exhibit bright outbursts with peak luminosities considerably
higher than $10^{37}$ \Lunit.  In 't Zand (2001) showed that during
the 2000 outburst, the source was active for at least 200
days. Moreover, our {\it Chandra} observations of the source were
performed $\sim$500 days after the start of the 2000 outburst and
$\sim$ 50 days before the 2001 re-flare, suggesting that in-between
those outburst episodes the source might have been active at similar
levels as detected during our {\it Chandra} observations (a few times
$10^{35}$ \Lunit). If indeed true, then the source was active for
almost 600 days and possibly even longer (no information is present
about the current state of this source). The bright 2000 and 2001
outbursts combined with the possible extended period of accretion are
in contrast with the basic properties of the faint X-ray transient
group (weak and short outbursts) making the classification of SAX
J1747.0--2853 as a faint X-ray transient no longer justified (unless
the classification criteria are relaxed).

A similar conclusion can be reached for the neutron star transient in
the globular cluster NGC 6440. This source exhibited a faint outburst
in 1998 August (In 't Zand et al. 1999), but it erupted again in 2001
August as a bright transient with a peak flux in excess of a few times
$10^{37}$ \Lunit~ and an outburst duration in excess of 3 months (In
't Zand et a. 2001). This behavior is remarkably similar to SAX
J1747.0--2853 in that a weak, short outburst was also followed by
brighter and much longer outbursts.  Those similarities between the
two systems raises the question if some (maybe all) of the other
systems classified as faint transients, can also exhibit bright
outbursts, which would cast doubts on whether those systems are indeed
fundamentally different from the bright X-ray transients. Such a
fundamental difference was suggested by King (2000) who argued that
those faint transients are binary neutron star systems which have
evolved below the period minimum, have low time-averaged mass
accretion rates ($\sim10^{-11}$ \mzon~year$^{-1}$), have recurrence
times of only a few years, and have very low-mass companion stars. The
large mass accretion rates inferred for SAX J1747.0--2853 and the
transient in NGC 6440 (In 't Zand et al. 2001) seem to rule-out such a
system configuration for these sources.  It also makes it less
unlikely that they will exhibit millisecond X-ray pulsations because
their relatively high inferred accretion rates will bury the magnetic
field of the neutron star (Cumming et al. 2001). Note, that the most
recent outbursts of NGC 6440 and SAX J1747.0--2853 might be atypical
for those sources and generally they might exhibit short and dim
outbursts, lowering their time-averaged accretion
rate\footnote{Similar arguments can also be applied to the normal
(i.e., bright) neutron star X-ray transients because of the lack of
knowledge about their long-term ($>1000$ years) X-ray behavior.}. Even
if true, the sources still cannot be classified as faint transients
(because their outbursts can be very bright and extended).  It remains
to be determined if a sub-group of faint X-ray transients exists and
if they are indeed post-minimum binaries.

\subsection{Low-level accretion activity}

We detected SAX J1747.0--2853 about 2 months before the 2001 outburst
peaked (which was around 17 September 2001), demonstrating that the
source was actively accreting well before the full outburst
happened. It demonstrates that SAX J1747.0--2853 has a complex
outburst behavior and it is possible that the source never returned to
quiescence ($<10^{34}$ \Lunit) after the 2000 outburst and stayed
active with typical luminosities of $10^{35}$ \Lunit. Such long-active
periods at low levels have been observed for different systems (e.g.,
the neutron star systems Aql X-1 and 4U 1608--52: Bradt et al. 2000;
$\check{{\rm S}}$imon 2002; Wachter et al. 2002; or the black-hole
system 4U 1630--47: Kuulkers et al. 1997).  The millisecond X-ray
pulsar SAX J1808.4--3658 also can exhibit episodes of long lived
low-level activity. Wijnands et al. (2001) found that the source was
active at a levels up to $\sim10^{35}$ \Lunit~for several months
during 2000. The source displayed luminosity swings up to three orders
of magnitudes on timescales of only a few days. It is unclear from our
observations if SAX J1747.0--2853 exhibited similar violent behavior
or that it was more stable, similar to what has been observed for the
other systems.  Our results add to the growing evidence that the
behavior of X-ray transients at low luminosity is very complex.

Several other systems have also been detected at similar low
luminosities. For example, Kaptein et al. (2000) and Cornelisse et al.
(2002a) detected, respectively, 1RXS J1718.2--402934 and SAX
J1828.5--1037 at luminosities of $10^{34-35}$ \Lunit. Cornelisse et
al. (2002a) suggested that they are part of the X-ray burst sources
which have low persistent luminosities ($10^{34-35}$ \Lunit) and might
form a separate sub-class of neutron-star X-ray binaries. But {\it
Chandra} observations of several group members showed that they could
only be seen at a luminosity of a few times $10^{32}$ \Lunit~ or less
(Cornelisse et al. 2002b). Those low luminosities are similar to the
quiescent luminosities observed for the normal neutron star X-ray
transients, suggesting that those low-luminosity persistent sources
might be genuine transients but could exhibit sub-luminous
($10^{34-35}$ \Lunit) outbursts episodes, possible with a durations of
years (see the discussion in Cornelisse et al. 2002b). Intriguingly,
SAX J1747.0--2853 likely displayed also an extended period (several
hundreds of days) of low-level accretion in-between the 2000 and 2001
outbursts. The possibility exists that the physical mechanism behind
this extended period of low-level activity in SAX J1747.0--2853 is
related to the mechanism behind the low-level activity in the low
persistent X-ray bursters. If true, it cannot be excluded that also
those latter source group might exhibit outbursts (either dim or
bright) in the future, similar to what has been observed for SAX
J1747.0--2853. Currently, it is unclear if different sub-classes of
neutron-star X-ray binaries are really needed to explain the observed
differences, or that the underlining mechanisms are related to each
other.

The traditional disk instability model to explain outburst light
curves of X-ray transients (see Lasota 2001 for an overview) has not
yet addressed the issue of the behavior of X-ray transients in this
low-luminosity region. The modeling of the behavior of neutron star
X-ray binaries in this regime might be complex because of the
increasing influence (see Campana et al. 1998a for a discussion) of
the magnetic field of the neutron star (if not buried completely by
the accreted matter). The lack of our knowledge is partly due to the
lack of sensitive observations in this regime.  However with current
instruments, we can now obtain high quality spectra of X-ray
transients when they are decaying into quiescence. SAX J1747.0--2853
was not the target of the {\it Chandra} observations reported here,
and therefore they were not optimized for the study of this object at
the detected brightness level. This resulted in relatively large
uncertainties in our spectral parameters which do not allow to make
strong conclusions.  The spectral fit results are consistent with a
constant spectral shape of the source between our observations and
that of the {\it BeppoSAX}/NFI observations reported by Natalucci et
al. (2000) when the source luminosity was approximately one order of
magnitude larger. No conclusive spectral softening or hardening was
observed, although we cannot exclude such spectral changes either.

\subsection{Future observations}

\begin{figure}
\begin{center}
\begin{tabular}{c}
\psfig{figure=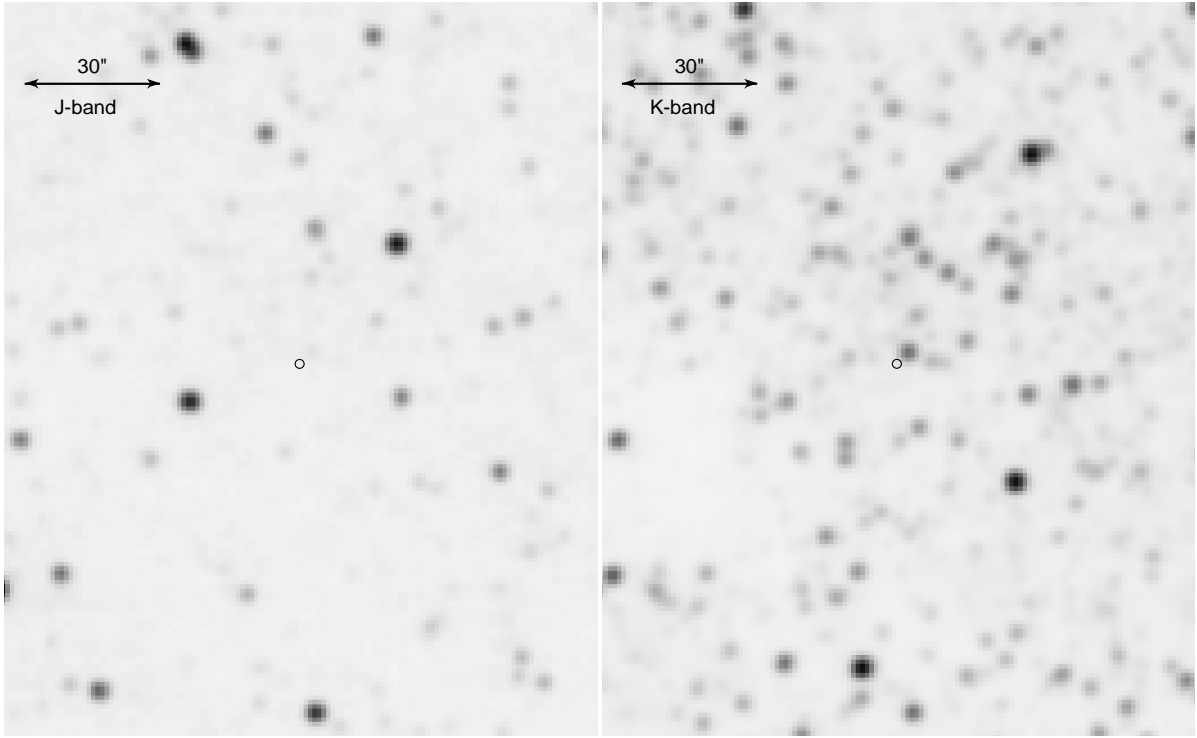,width=16cm}
\end{tabular}
\figcaption{The J-band (left) and $K_s$-band finding charts of SAX
J1747.0--2583. North is up and East is left. The solid circle is the
{\it Chandra} error circle of SAX J1747.0--2853.
\label{fig:finding} }
\end{center}
\end{figure}

The sub-arcsecond {\it Chandra} position of SAX J1747.0--2853 allows
for follow up studies at other wavelengths at times when the source is
found to be active. Although the high column density of the source
will strongly inhibit detection of the source at optical wavelengths,
it might be possible to detect the source at (near-)infrared
wavelengths. To this order we have presented $J$ and $K_s$ finding
charts of SAX J1747.0--2853 in Figure~\ref{fig:finding}. The excellent
position of the source might also be useful for follow up X-ray
observations with {\it Chandra} or {\it XMM-Newton} when the source is
in quiescence. Quiescent neutron star systems have luminosities in the
range of $10^{32}$ to $10^{33}$ \Lunit, resulting in a flux range of
$10^{-14}$ to $10^{-13}$ \funit. If the quiescent flux of SAX
J1747.0--2853 is dominated by the same thermal component observed in
other quiescent neutron star systems (with blackbody temperature of
0.2--0.3 keV; e.g., Bildsten \& Rutledge 2002 and references therein),
then it is unlikely (due to the large column density) that the source
will be detectable within a reasonable amount of time (less than a few
tens of kiloseconds) with either {\it Chandra} or {\it
XMM-Newton}. However, if besides the thermal component, also a
power-law component above 2 keV is present (as observed for several
quiescent systems; e.g., Asai et al. 1996, 1998; Campana et al. 1998b)
and it contributes to about half the total flux, than with a few tens
of kiloseconds the source might be detectable (similarly to the recent
detection of quiescent emission from the neutron star X-ray transient
GRO J1744--28, which has a comparable column density; Wijnands \& Wang
2002). Due to the sub-arcsecond {\it Chandra} position, the quiescent
counterpart of SAX J1747.0--2853 can be easily identified. For other
transients near the Galactic center region, the positions are usually
known only to an accuracy of 1 arcminutes or worse, which increases
significantly the probability that if any sources are detected in the
error circles of those transients, that they are background AGNs or
unrelated Galactic sources which also emit hard X-rays.

\acknowledgments

RW was supported by NASA through Chandra Postdoctoral Fellowship grant
number PF9-10010 awarded by CXC, which is operated by SAO for NASA
under contract NAS8-39073. WQD is supported by the CXC grant
SAO-GO1-2150A. This publication makes use of data products from the
Two Micron All Sky Survey, which is a joint project of the University
of Massachusetts and the Infrared Processing and Analysis
Center/California Institute of Technology, funded by the National
Aeronautics and Space Administration and the National Science
Foundation. This publication also used the quick-look results provided
by the ASM/{\it RXTE} team and resources provided through the HEASARC
on-line service, provided by the NASA-GSFC. We thank John Houck and
the CXC ISIS team for useful discusses about ISIS.

\begin{figure}
\begin{center}
\begin{tabular}{c}
\psfig{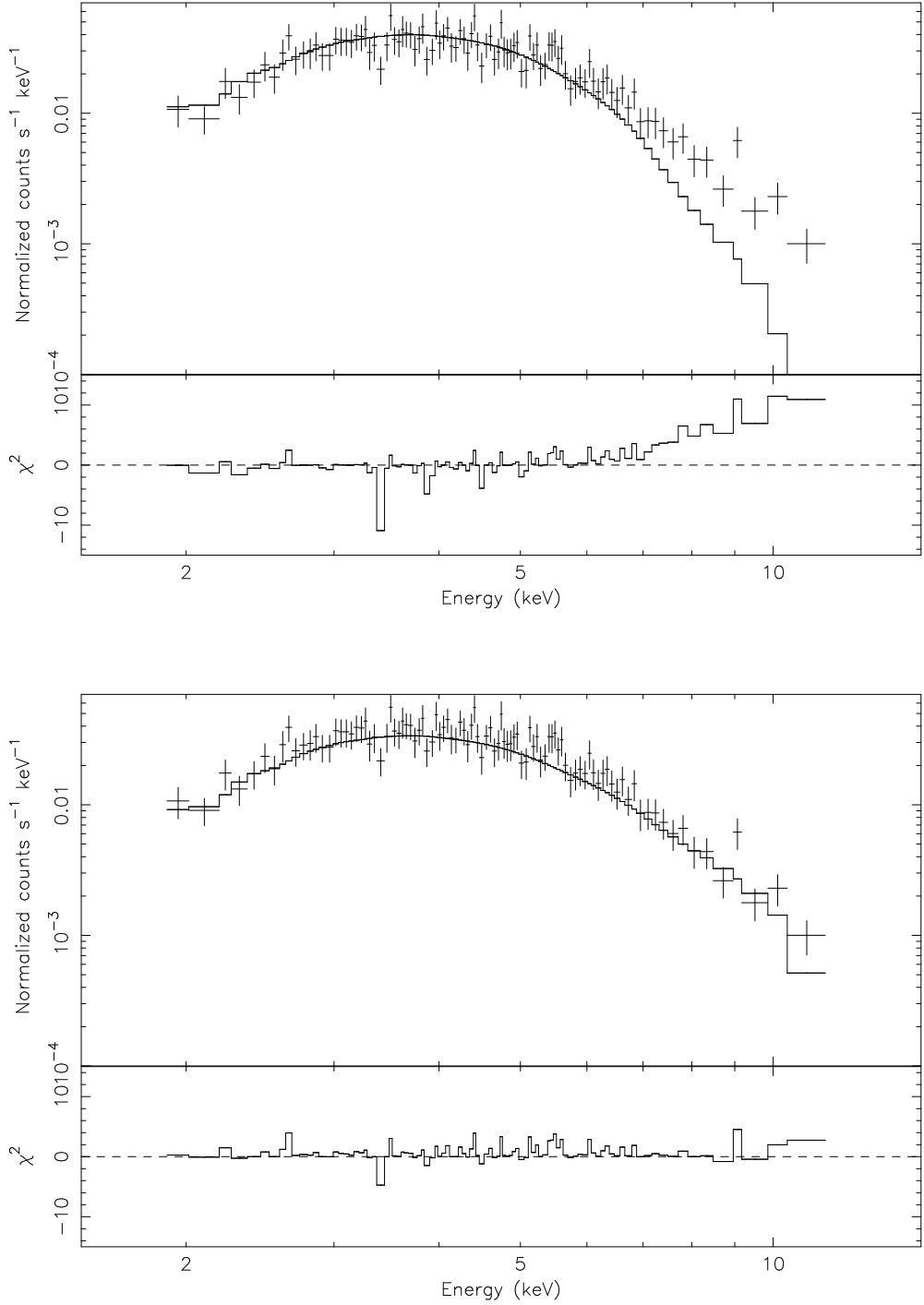}
\end{tabular}
\figcaption{The {\it Chandra}/ACIS-I spectrum for SAX J1747.0--2853 as
obtained during observation GCS 10. The solid line in the top panel is
the best fit power-law as obtained using ISIS and the pile-up model,
but no pile-up correction was applied for display purposes. Clearly,
excess counts are present for photon energies above 6 keV. The solid
line in the bottom panel shows the same fit, but now also the pile-up
correction has been applied. Clearly, the power-law + pile-up model
produces an acceptable fit to the data.\label{fig:spectrum} }
\end{center}
\end{figure}

\begin{figure}
\begin{center}
\begin{tabular}{c}
\psfig{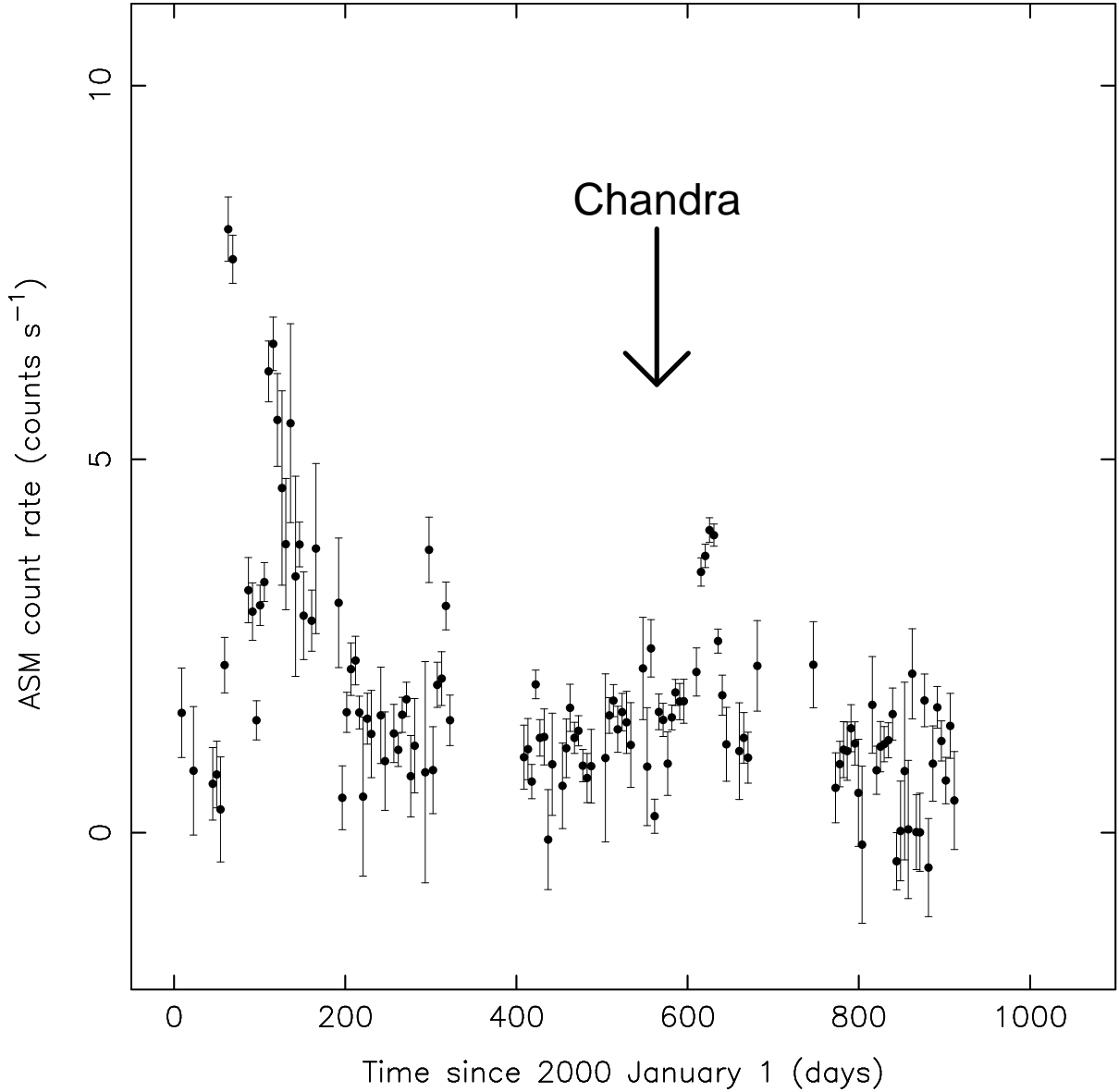}
\end{tabular}
\figcaption{The {\it RXTE}/ASM light curve of SAX J1747.0--2853 since
2000 January 1 in 5 day bins. The arrow indicates at which time the
{\it Chandra} observations were taken.  Although it appears that the
source was weakly detected by the {\it RXTE}/ASM at the time of the
{\it Chandra} observations, this is due to the uncertainties in the
fitting process to obtain the source counts (Levine et al. 1996) and
the proximity of the source to the Galactic center with introduces a
considerably amount of extra noise.
\label{fig:asm_lc} }
\end{center}
\end{figure}

\begin{deluxetable}{lcccccl}
\tablecolumns{7}
\tablecaption{{\itshape Chandra} observations of SAX J1747.0--2853
\label{tab:log}}
\tablehead{
Obs-ID & Start date     & End date          & Exposure   & Off-axis$^a$& Chip$^b$ & Comments \\
       & \multicolumn{2}{c}{(2001 July 18)} &  (ksec)    & ($'$)       &          &}
\startdata
GCS 10  &  00:48         & 04:17             & $\sim$11.6 & $\sim$2.7  & ACIS-I2   &\\
GCS 11  &  04:17         & 07:45             & $\sim$11.6 & $\sim$14.3 & ACIS-S2   & At chip edge \\
\enddata

\tablenotetext{a}{The distance of the source to the pointing direction.}

\tablenotetext{b}{The chip on which the source was located.}

\end{deluxetable}

\begin{deluxetable}{lcccl}
\tablecolumns{5}
\tablecaption{Spectral fit results of SAX J1747.0--2853$^a$ \label{tab:spectrum}}
\tablehead{
Obs-ID & $N_{\rm H}$           & Photon index        & Flux$^b$            & Comments\\
       & ($10^{22}$ cm$^{-2}$) &                     & ($10^{-11}$ \funit) &}
\startdata
GCS 10 & 7\pp1                 & 1.8$^{+0.3}_{-0.4}$ & 3.3                 & Pile-up-corrected\\
       & 7$^{+3}_{-2}$         & 2.0$^{+1.0}_{-0.8}$ & 3.6                 & Annulus extraction region\\
GCS 11 & 6.9$^{+1.5}_{-0.9}$   & 2.1\pp0.5           &                     & Source at edge of S2 chip \\
\enddata

\tablenotetext{a}{The errors on the fit parameters are for 90\%
confidence levels.}

\tablenotetext{b}{The fluxes are unabsorbed and for 0.5--10 keV. The
fluxes listed for the second row of GCS10 are corrected for the
annulus extraction region. No flux is listed for GCS11 because it is
unclear what the exposure correction factor is (see text).}

\end{deluxetable}

\end{document}